\documentclass[10pt,conference]{IEEEtran}

\usepackage{cite}

\ifCLASSINFOpdf
   \usepackage[pdftex]{graphicx}
\else
   \usepackage[dvips]{graphicx}
\fi
\usepackage{amsmath}
\usepackage{url}

\begin{document}
%
\title{Release as a Contract: A Concept of Meta-Maintenance for the Entire FLOSS Ecosystem}

\author{
\IEEEauthorblockN{Hideaki Hata}
\IEEEauthorblockA{\textit{Shinshu University} \\
hata@shinshu-u.ac.jp}
}


%


\maketitle

\begin{abstract}
We advocate for a paradigm shift in supporting free/libre and open source software (FLOSS) ecosystem maintenance, from focusing on individual projects to monitoring a whole organic system of the entire FLOSS ecosystem, which we call software \textit{meta-maintenance}. We discuss challenges of building a global source code management system, a global issue management system, and FLOSS human capital index, 
based on the blockchain technologies.
\end{abstract}


%
\IEEEpeerreviewmaketitle

\section{Introduction and The Vision}

Free/libre and open source software (FLOSS) is characterized as a privately produced public
good and a common pool resource (non-exclusive)~\cite{OMAHONY20031179}.
Based on various internal and external motivations, developers engage in FLOSS~\cite{Coelho:2018:WWE:3195836.3195848}, and FLOSS projects have a tremendous impact on our daily lives.

However, because of the nature of free and open source, and the lack of the view of the whole picture, there are several problems even in the aspect of code.
\textit{(i) Broken library dependencies}.
Even though third-party library reuse is common practice, updating libraries is not as common.
From the study of library migration in more than 4,600 \texttt{GitHub} software projects,
Kula et al. reported that 81.5\% of the studied software used outdated libraries~\cite{Kula2017}. From the survey, they also found that 69\% of the answered developers claimed to be unaware of their vulnerable dependencies.
\textit{(ii) License violations}.
From the study of the Android application market \texttt{F-Droid}, Mlouki et al. found license violations in 17 out of 857 applications~\cite{7476659}. They also reported that many files are not licensed in their first release.
\textit{(iii) Ad hoc code reuses}.
Copying software components and then maintaining them by a new owner is one type of code reuse, which is known as \textit{clone-and-own}~\cite{Dubinsky:2013:ESC:2495256.2495759,Ishio:2017:SFS:3104188.3104222}.
In software product line engineering, the usage of this clone-and-own approach is discouraged, since it makes maintaining multiple product lines difficult. If changes are made in the original or copied components, they are not easily propagated. This is because of the ad hoc nature of component reusing. Developers working in different product lines do not know when clone-and-own operations occurred and where the cloned components are located. Although there are disadvantages, the clone-and-own approach is adopted in FLOSS projects as well as industrial software product lines due to its benefits, such as simplicity, availability, and independence of developers~\cite{Dubinsky:2013:ESC:2495256.2495759}.

Even with its large-scale code and a large amount of developers, Google do not have such issues~\cite{Potvin:2016:WGS:2963119.2854146}. As Google is a single company, it is not questionable not to have license issues. Then how do they solve other problems? Google is reported to have a monolithic source code management system, which have several advantages: unified versioning, extensive code sharing and reuse, simplified dependency management, atomic changes, large-scale refactoring, collaboration across teams, flexible team boundaries and code ownership, and code visibility and clear tree structure providing implicit team namespacing~\cite{Potvin:2016:WGS:2963119.2854146}.
Based on the monolithic source code management, changes to core libraries are promptly and easily propagated through the dependency chain into the final products that rely on the libraries~\cite{Potvin:2016:WGS:2963119.2854146}.
How can we realize this scenario in FLOSS development?

We advocate for a new vision of supporting FLOSS maintenance, \textbf{meta-maintenance}, which intends to monitor and support a whole organic system of the entire FLOSS projects.
In our previous research, we analyzed the potential of meta-maintenance targeting the maintenance of common source code files in multiple repositories~\cite{10.1109/ICSE43902.2021.00076}, but in this article we will focus on the concept of maintaining the entire ecosystem, that is,
we intend to introduce Google-like code and knowledge sharing systems, not in within a single organization, but for the whole FLOSS projects.
Because contributors voluntarily work to individual FLOSS projects and they do not forfeit their rights to it, we cannot have the same systems with large-scale single companies like Google.
To maintain source code, track issues and resources, monitor project health, and recommend sharing knowledge, we discuss the following research challenges that need to be addressed for meta-maintenance.

\begin{itemize}
\item \textbf{Global source code management system.} To address code-related problems in FLOSS, we propose a concept of a system of globally shared single source of truth, a \textit{global version control system}, instead of a monolithic version control system.
\item \textbf{Global issue management system.} It is reported that, in Google, technical debt caused by dependent systems is paid down immediately as changes are made~\cite{Potvin:2016:WGS:2963119.2854146}. To track such cross-project (product) issues, we consider a \textit{global issue tracking system} is necessary.
\item \textbf{FLOSS human capital index.} FLOSS projects can be no longer under maintenance because of several reasons and cause negative impacts to the client projects~\cite{Coelho:2017:WMO:3106237.3106246}.
Similar to the Global Human Capital Index (GHCI)\footnote{\url{http://reports.weforum.org/global-human-capital-report-2017/}} proposed and reported by the World Economic Forum, we need a system for monitoring individual and the whole FLOSS project health in terms of human capital~\cite{WOHLIN-2015,2018arXiv180503844O}.
\end{itemize}

Challenges include not only building individual systems but also integrating those systems effectively.
The main idea of building global systems is adopting blockchain technologies.


\section{Preliminary}

\subsection{Blockchain}
A blockchain is a chronologically irreversible distributed ledger that records history of transactions between nodes in a community.
Blockchain was originally proposed as a solution to the double-spending problem of electronic payments by Satoshi Nakamoto in 2008~\cite{nakamoto2008bitcoin}.

To elucidate the characteristics of blockchain, we can compare it with a traditional client-server model: 
(i) each node behaves as a predefined roll, client or server, 
(ii) all requests from clients are executed through a server,  
(iii) all history of transactions is stored only in a server. 
In contrast, a blockchain is based a peer-to-peer (P2P) model, which has the different characteristics:
(i) every node is equivalent to each other and have a fair authority,
(ii) every transaction is directly executed between two peers,
(iii) all history of transactions is stored in a distributed ledger that all nodes share.



\begin{figure}
\includegraphics[width=\linewidth]{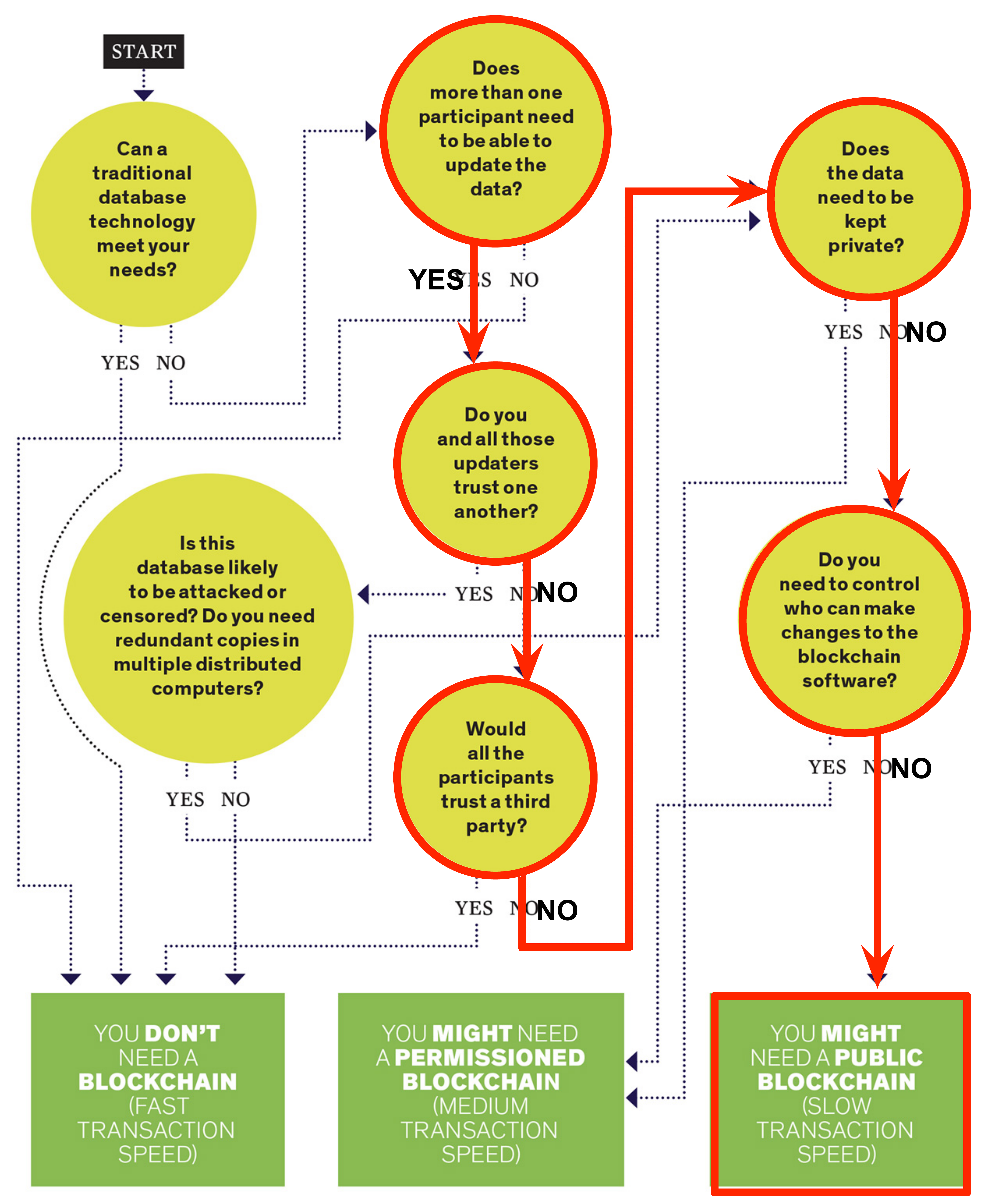}
\caption{Do you need a blockchain? Chart from~\cite{8048838}. We emphasize one flow.}
\label{fig:chart}
\end{figure}

Our idea of building global systems (a global version control system and a global issue tracking system) is recording all \textit{valid} transactions in a public blockchain, which allows participants in the whole FLOSS ecosystem to reliably know how and who conduct transactions, and also enable them to validate those transactions.
When considering versioning systems, we have already had an established modern version control system, \texttt{Git}. However, we consider that traditional version control systems does not meet the needs.
In order to clarify the problems, we use the flowchart presented in the article~\cite{8048838}, and shown in Figure~\ref{fig:chart}.

\textbf{Does more than one participant need to be able to update the data? Yes.}
Since anyone can make clone-and-own reuses, anyone can be the participants in the system. If we will use a single \texttt{Git} repository as a global versioning system, obviously it is not a good idea to make all participants committers, who can record versioning transactions. The pull-based model does not scale because it requires integrators~\cite{Gousios:2015:WPC:2818754.2818800}.
As studied in the previous work, Android project manages their own Git repositories for cloned components and records histories that can match for the original component versions~\cite{Ishio:2017:SFS:3104188.3104222}. Although Git can work in a single organization, it should be difficult to make a single Git repository work in many organizations.
Among various software components and various participants involved in clone-and-own transactions, we assume no one can be integrators to validate all reuse transactions to be recorded in the system.

\textbf{Do all those updaters trust one another? No.}
As stated before, trusting all participants and making them comitters cannot work. This is not only because of possible evil participants who are trying to record false information. Even ordinary participants can record invalid clone-and-own transactions, for example, license violated reuses. As reported in the previous study, developers seem to have some difficulties in understanding licenses~\cite{7476659}.

\textbf{Would all the participants trust a third party? No.}
From the above discussion, having a single (or even having multiple and integrated) \texttt{Git} repository seems not to be a good solution.
If we will have an alternative and more suitable version control database, locating it in a hosting service like \texttt{GitHub} can be an option. However, we consider this is also not good. If this hosting service goes down, we cannot record and investigate clone-and-own transactions. This is worse than the down of \texttt{GitHub}, because we do not have a local database.
Essentially, granting someone or a third party special permission seems to be inappropriate for this global system.

\textbf{Does the data need to be kept private? No.}
Clone-and-own data should be public. Transparent processes and records should be desirable.

\textbf{Do you need to control who can make changes to the block-chain software? No.}
Again, anyone should not have special permission nor be an admin.

In summary, we cannot use \texttt{Git} for a global versioning system, nor adopt a pull-based model. We should not have cloud servers since they can be a point of failure that can stop the entire system. We need a public and transparent recording system. As the flowchart in Figure~\ref{fig:chart} shows, we might \textit{need a public blockchain.}
We can have similar discussions to a global issue tracking system compared with traditional systems.

\subsection{Smart Contracts}

Nick Szabo introduced the concept of a smart contract as \textit{a computerized transaction protocol that executes the terms of a contract}~\cite{szabo1994sc}.
Within the blockchain context, smart contracts are scripts stored on the blockchain~\cite{7467408}.
By addressing a transaction, we can trigger a smart contract. It then executes independently and automatically in a prescribed manner on every node in the blockchain network. Smart contracts operate as autonomous actors, whose behaviors are completely predictable.

The most prominent framework for smart contracts is considered to be Ethereum~\cite{Atzei:2017:SAE:3080353.3080363}, a blockchain platform that remotely executes software on a distributed computer system called the Ethereum Virtual Machine.
Ethereum smart contracts generally serve four purposes~\cite{ethereumDT}.
\begin{itemize}
\item Maintain a data store. For example, simulating a currency and recording membership in a particular organization.
\item Manage an ongoing contract or relationship between multiple users. One example of this is a contract that automatically pays a bounty to whoever submits a valid solution to some mathematical problem
\item Resend incoming messages to some desired destination only if certain conditions are met. This is called a forwarding contract.
\item Provide functions to other contracts like a software library.
\end{itemize}

\subsection{Distributed Autonomous Organizations}

Based on smart contracts, there is a concept of ``decentralized autonomous organizations'' (DAOs)~\cite{7467408}.
The organization members follow rules written in smart contracts, and those rules can be modified based on the voting of members. With our concept of global systems, all participants can be regarded as members of a DAO. Everyone, including smart agents, share information, follow rules and manage rules, based on autonomous and reliable mechanisms.

\section{Challenges}

\subsection{Global Version Control System}

\begin{figure}
\includegraphics[width=\linewidth]{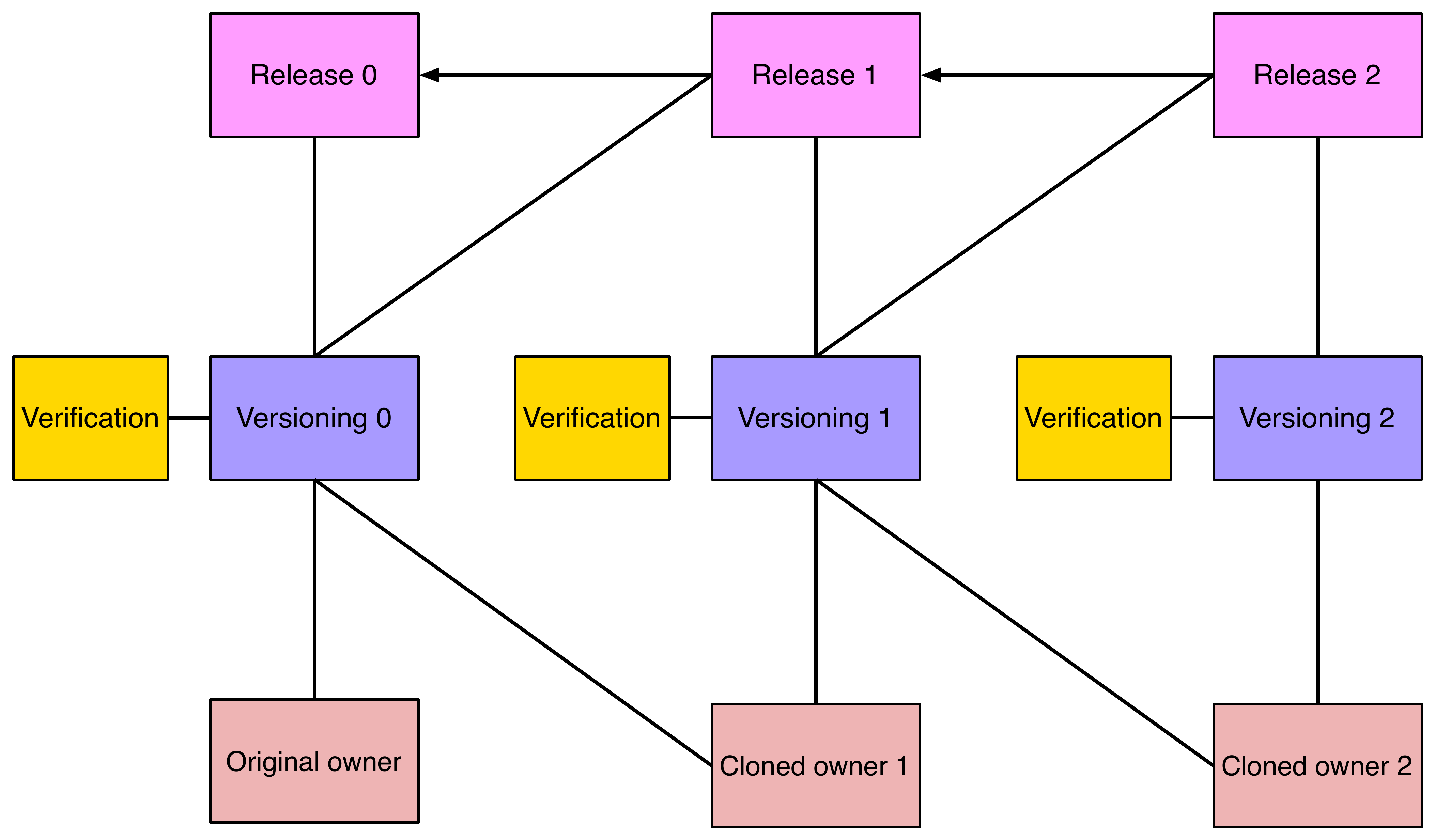}
\caption{A temporary idea of the overview of release as a contract. The relationships of smart contracts among the original owner and cloned owners in a case of sequential clone-and-own reuses. Forwarded contracts are ignored for simplicity.}
\label{fig:idea}
\end{figure}

A blockchain enables all stakeholders to have one shared database to keep track of, where updates from any participants come with cryptographic verifiability, get propagated along the blockchain network automatically.
Similarly, we can keep track of reused components with a shared database, a blockchain network.
When participants operate transactions like shippin new releases, they can create smart contracts or send signed message to predetermined and agreed-upon smart contracts to allow every participant on the blockchain network to know them.
%
We can allow any participants to update data without a third party. All participants do not need to trust one another. Data is public and no specific single participant can control the system.

Designing smart contracts is the core challenge of the system. We introduce a concept of \textit{release as a contract}. 
As shown in Figure~\ref{fig:idea}, when the original owner organization releases a reusable software component, they create the release 0 contract, which describe names and versions of the component. If there are updates, newer versions are written in the contract via the versioning 0 contract, which is also prepared by the original owner. If another team (cloned owner 1) reuses the component and release their software with the cloned component, they send messages to the versioning 0 contract to notify this clone-and-own transaction. The cloned component is verified with prepared libraries whether it satisfies the license and other rules made by the original owner. Only if this verification succeeds, the cloned owner 1 is allowed to create the release 1 contract that refers the specific version of the original component written in the release 0 contract. Similarly, cloned owner 2 will create the release 2 contract after passing the verification. If there are update changes in upstream releases, they can be propagated with the tracking information.

As we discussed previously, smart contracts operate autonomously based on predetermined programs. Versions of components from the same owners are written in the same release contracts, and the clone-and-own relationships are seen in the references. These data are all public and the recording process is transparent. In addition, there is no human integrator in this versioning process. Since a smart contract can be prepared for each release, owner can determine rules of reuses. If some participants cannot agree on a contract, they do not need to sign the contract, or they can negotiate with the owner to change the contract.

If users prefer to use software components released on blockchain as contracts, isolating from this autonomous ecosystem will not be a good choice for development organizations. If we will develop efficient and accurate verification tools for licenses and hidden copy detection, lying should be difficult and human errors can be decreased.

Note that we do not intend to replace all Git repositories with the blockchain system. Instead, our global version control system consists of local independent version control systems like Git and one blockchain that only record reuse transactions.
We have conducted an empirical analysis of the clone-and-own reuses exist in the FLOSS ecosystem~\cite{2022arXiv220406531H}.

\subsection{Global Issue/Resource Tracking System}

Similar to the above global version control system, we can consider \textit{issue report as a contract}.
Versions of dependent components/products, description of the issues, affected products can be associated with smart contracts of issues. To make the system practical, issue searching and mechanism to incentivize contributors to work on cross-project issues are needed to be developed.

\subsection{FLOSS Human Capital Index}

In economics, human investments are seen as capital -- specifically intellectual capital.
Oxford handbook on Human Capital highlights the importance of human capital, that is,
\textit{``all forms of intellectual capital including social and structural capital are arguably reducible to the human knower, thus human capital becoming the linchpin''}~\cite{RePEc:oxp:obooks:9780199655892}.
A practical implementation of human capital is the Global Human Capital Index (GHCI)\footnote{http://reports.weforum.org/global-human-capital-report-2017/}, as proposed and reported by the World Economic Forum.
As previous study started~\cite{WOHLIN-2015,2018arXiv180503844O},
building such human capital index for FLOSS is another challenge.



\section{Conclusion}
This paper introduces a concept of meta-maintenance for the whole FLOSS projects.
mainly using blockchain and smart contract technologies.
Based on the preferable P2P architecture of blockchain, we design the smart contract relationship for the idea of release as a contract and issue as a contract. Building human capital index for health indicator 
is another challenge. Obviously there are many challenges technically and socially. The initial step of this work should be building a working prototype.






\bibliographystyle{IEEEtran}
\bibliography{ref}
%

\end{document}